\documentclass[twocolumn]{aastex62}

\usepackage[caption=false]{subfig}
\usepackage{float}
\usepackage{graphicx}	
\usepackage{amsmath}	
\usepackage{amsfonts}	
\usepackage{amssymb}
\usepackage{xcolor}
\usepackage{hhline}
\usepackage{bm}
\usepackage[version=3]{mhchem}

\newcommand{\mpl}{\ensuremath{M_{p}}}
\newcommand{\mjup}{\ensuremath{M_{\rm J}}}
\newcommand{\kms}{\ensuremath{\rm km\,s^{-1}}}

\newcommand{\obliq}{\ensuremath{107.9^{+2.0}_{-1.7}}}
\newcommand{\obliqwithunits}{\ensuremath{{107.9^\circ}^{+2.0^\circ}_{-1.7^\circ}}}

\shorttitle{A profusion of species in HAT-P-70\,b}
\shortauthors{Bello-Arufe et al.}
\graphicspath{{./}{figures/}}

\begin{document}

\title{Mining the Ultra-Hot Skies of HAT-P-70b: Detection of a Profusion of Neutral and Ionized Species}

\correspondingauthor{Aaron Bello-Arufe}
\email{aarb@space.dtu.dk}

\author[0000-0003-3355-1223]{Aaron Bello-Arufe}
\affil{National Space Institute, Technical University of Denmark, Elektrovej, DK-2800 Kgs. Lyngby, Denmark}

\author[0000-0001-9749-6150]{Samuel H. C. Cabot}
\affil{Yale University, 52 Hillhouse Avenue, New Haven, CT 06511, USA}

\author[0000-0002-6907-4476]{João M. Mendonça}
\affil{National Space Institute, Technical University of Denmark, Elektrovej, DK-2800 Kgs. Lyngby, Denmark}

\author[0000-0003-1605-5666]{Lars A. Buchhave}
\affil{National Space Institute, Technical University of Denmark, Elektrovej, DK-2800 Kgs. Lyngby, Denmark}

\author[0000-0002-4227-4953]{Alexander D. Rathcke}
\affil{National Space Institute, Technical University of Denmark, Elektrovej, DK-2800 Kgs. Lyngby, Denmark}

\begin{abstract}

With an equilibrium temperature above 2500 K, the recently discovered HAT-P-70\,b belongs to a new class of exoplanets known as ultra-hot Jupiters: extremely irradiated gas giants with day-side temperatures that resemble those found in stars. These ultra-hot Jupiters are among the most amenable targets for follow-up atmospheric characterization through transmission spectroscopy. Here, we present the first analysis of the transmission spectrum of HAT-P-70 b using high-resolution data from the HARPS-N spectrograph of a single transit event. We use a cross-correlation analysis and transmission spectroscopy to look for atomic and molecular species in the planetary atmosphere. We detect absorption by \ion{Ca}{2}, \ion{Cr}{1}, \ion{Cr}{2}, \ion{Fe}{1}, \ion{Fe}{2}, \ion{H}{1}, \ion{Mg}{1}, \ion{Na}{1} and \ion{V}{1}, and we find tentative evidence of \ion{Ca}{1} and \ion{Ti}{2}. Overall, these signals appear blue-shifted by a few km\,s$^{-1}$, suggestive of winds flowing at high velocity from the day-side to the night-side. We individually resolve the \ion{Ca}{2} H \& K lines, the \ion{Na}{1} doublet, and the H$\alpha$, H$\beta$ and H$\gamma$ Balmer lines. The cores of the \ion{Ca}{2} and \ion{H}{1} lines form well above the continuum, indicating the existence of an extended envelope. We refine the obliquity of this highly misaligned planet to $107.9^{+2.0}_{-1.7}$ degrees by examining the Doppler shadow that the planet casts on its A-type host star. These results place HAT-P-70\,b as one of the exoplanets with the highest number of species detected in its atmosphere.
\end{abstract}



\section{Introduction} \label{sec:intro}
Ultra-hot Jupiters constitute the hottest class of exoplanets. The day-sides of these extremely irradiated gas giants are characterized by temperatures well above $2000~ \rm{K}$, reminiscent of those found in stars and hot enough to dissociate most molecular species \citep{arcangeli2018,bell2018,parmentier2018,lothringer2018}. Their high temperatures provide them with inflated atmospheres and a day-side and evening terminator that are mostly cloud-free and near chemical equilibrium \citep{heng2016,kitzmann2018,gao2020,helling2021}. These exotic worlds are therefore ideal targets for atmospheric characterization through transmission spectroscopy.

Many of the current efforts to characterize ultra-hot Jupiters involve ground-based observations with high-resolution spectrographs. These observations are revealing a large diversity of neutral and ionized atomic metals in ultra-hot Jupiters \citep[e.g.][]{hoeijmakers2019,ben-yami2020}. The study of these refractory elements in transmission spectroscopy can inform us about the chemistry at the terminator region, the speed of the winds flowing between the day-side and the night-side, and potentially the formation and migration history of the planet \citep[e.g.][]{ehrenreich2020,nugroho2020,lothringer2020,borsa2021,kesseli2021}.

HAT-P-70\,b is a recently discovered ultra-hot Jupiter \citep{zhou2019}. It has a mass of $\mpl<6.78\,\mjup$, and it transits an A-type star with $T_{\textup{eff}} = 8450^{+540}_{-690}~\textrm{K}$ every 2.74 days. At a distance of only 0.047 AU from its host star, HAT-P-70\,b is subject to extreme stellar irradiation that elevates the equilibrium temperature of the exoplanet to $T_{\textup{eq}} = 2562^{+43}_{-52}~\textrm{K}$. Its large radius ($R_p=1.87~\text{R}_\text{J}$) and the brightness of its host star ($V=9.47$) make HAT-P-70\,b a very favorable target for follow-up high-resolution transmission spectroscopy. 

In this work we present the first analysis of the transmission spectrum of HAT-P-70\,b. We analyze high-resolution observations from a single transit event observed with HARPS-N to search for absorption by atomic and molecular species via transmission spectroscopy and the cross-correlation technique \citep{snellen2010}. This paper is structured in the following way: Section~\ref{sec:observations} describes the observations, data reduction and telluric correction. Section~\ref{sec:crosscorr} introduces our transmission spectroscopy and cross-correlation analyses, and how the Doppler shadow that results from the cross-correlation analysis is useful to refine the obliquity of HAT-P-70\,b. Section~\ref{sec:resultsdiscussion} reports and discusses the results, and Section~\ref{sec:conclusions} presents the conclusions.

\section{Observations and data reduction} \label{sec:observations}

HARPS-N is a high-resolution ($R \sim 115,000$), fiber-fed, cross-dispersed echelle spectrograph covering the wavelength range between 3830\,\AA\ and 6930\,\AA\ \citep{cosentino2012}. It is mounted on the 3.58 m Telescopio Nazionale Galileo (TNG) at the Observatorio del Roque de los Muchachos on the island of La Palma, Spain. Using HARPS-N, we observed a transit of HAT-P-70\,b on the night of 18 Dec 2020 between 21:28 and 02:10 UTC (program: A42TAC27, PI Bello-Arufe). Although we were scheduled to observe until 03:00 UTC, technical issues related to the tracking of the telescope caused the observations to stop earlier. Nevertheless, we still captured the full transit event, with sufficient out-of-transit baseline before and after the transit. We set the exposure time to 400 seconds, obtaining 40 exposures, 30 of which during transit. Fiber A was placed on the target and fiber B on the sky. The night was clear with an average seeing of 0\farcs9--1\farcs0. The airmasses of the different exposures were in the range 1.36--1.06--1.20.

The data were reduced using the HARPS-N Data Reduction Software (DRS), version 3.7. The main steps include bias and dark subtraction, bad pixels and cosmics correction, flat fielding, and wavelength calibration. Although the DRS also provides 1-dimensional extracted spectra where the different orders are merged and rebinned, we only used these 1d spectra to produce the telluric model. In the rest of our analysis we used the 2-dimensional (i.e. order by order) extracted spectra to avoid artifacts of order stitching. The signal-to-noise ratio (S/N) per pixel as measured by the DRS at the center of order 64 (the order with H$\alpha$) ranges from 28.5 to 51.2.

\subsection{Telluric Correction}
Gases in the atmosphere of the Earth produce absorption lines in spectra taken from ground-based observatories. In particular, the wavelength range covered by HARPS-N is affected by absorption predominantly due to oxygen and water. We used the software package \texttt{molecfit} version 1.5.9 \citep{smette2015,kausch2015} to correct for these telluric absorption lines. \texttt{Molecfit} combines an atmospheric profile from the Global Data Assimilation System (GDAS) website and the line-by-line radiative transfer code LBLRTM \citep{clough2005} to fit user-selected regions of the observed spectra and produce a synthetic transmission spectrum of the atmosphere of the Earth. \texttt{Molecfit} is frequently applied to HARPS-N data \citep[e.g.][]{hoeijmakers2018,casasayasbarris2019,wyttenbach2020,stangret2021}, and it is a more robust way to correct for telluric absorption than other empirical methods, e.g. using airmass \citep{langeveld2021}.

\begin{figure*}
    \centering
    \includegraphics[width=\linewidth]{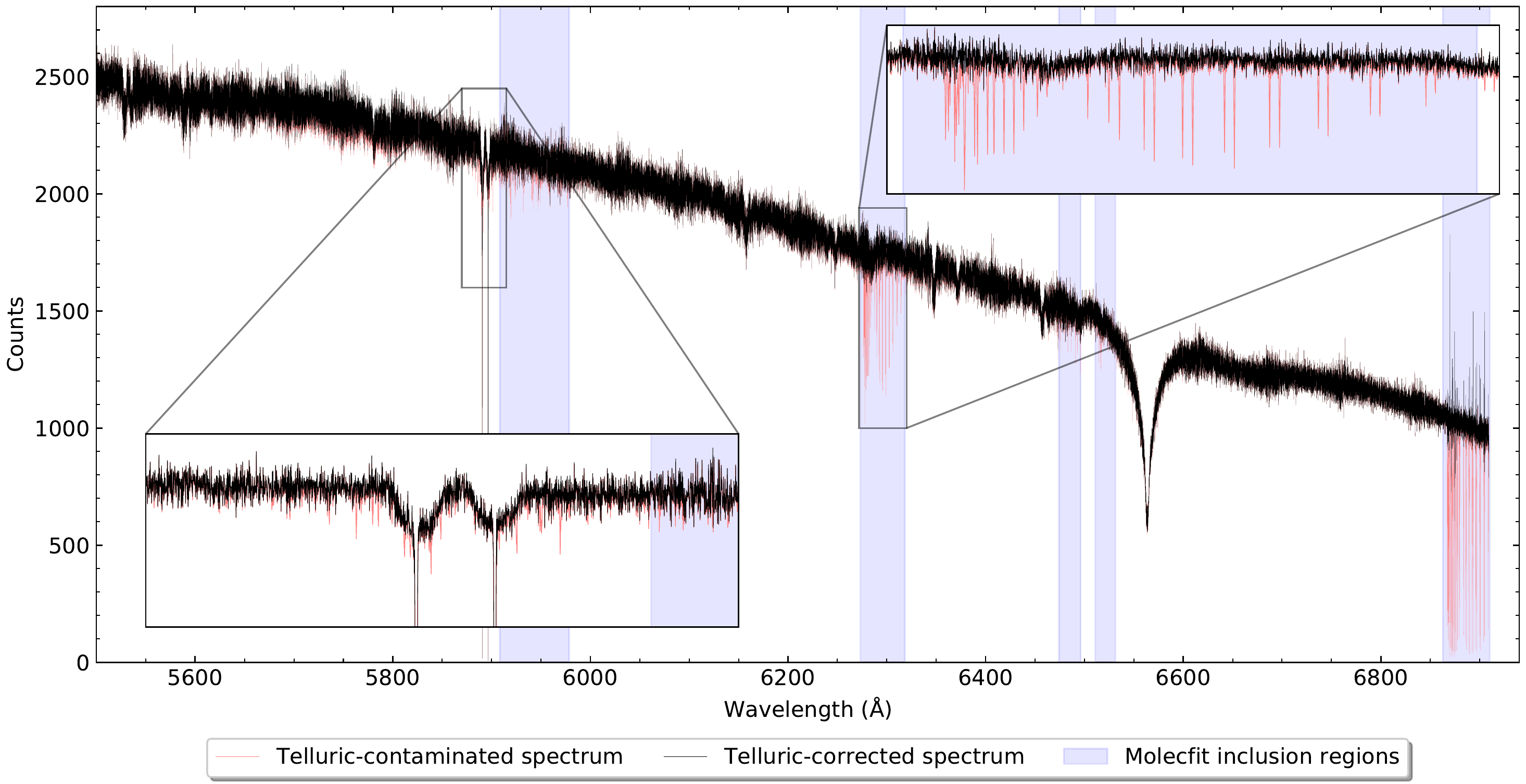}
    \caption{One of the HARPS-N spectra of HAT-P-70 before (red) and after (black) correcting the telluric absorption with \texttt{molecfit}. Only the red part of the spectrum is shown, where most telluric lines are found. The regions included in the telluric fit are highlighted in blue. The two insets show in more detail the spectral regions around the \ion{Na}{1} doublet (left) and the oxygen $\gamma$-band (right). Deep absorption is visible in the core of the stellar \ion{Na}{1} doublet, which we attribute to interstellar medium absorption.}
    \label{fig:molecfit}
\end{figure*}

We ran \texttt{molecfit} on the 1-dimensional extracted spectra after Doppler-shifting them from the rest frame of the Solar System barycenter to that of the observer. The regions that were included in the telluric fit were selected such that they covered a flat continuum with water and oxygen telluric features and without significant stellar absorption features. We fit a third-degree polynomial to the continuum of each spectral region. Given the high stability of HARPS-N, we did not re-calibrate the wavelength solutions of the different spectra. We found that values of $10^{-5}$ for both the relative $\chi^2$ and parameter convergence criteria improved the quality of the fit. The telluric models produced by \texttt{molecfit} for each 1-dimensional spectrum are then used to correct the 2-dimensional spectra via linear interpolation. Figure~\ref{fig:molecfit} shows one of the 1-dimensional spectra before and after telluric correction, including the spectral regions that were part of the fit.

\section{Methods} \label{sec:crosscorr}
In this section, we describe the methods we used to search for atomic and molecular species in the atmosphere of HAT-P-70\,b and how we derived the obliquity of the planet using its Doppler shadow. 

\subsection{The Cross-Correlation Analysis}\label{sec:metalsearch}
Since the discovery of carbon monoxide in the atmosphere of the hot Jupiter HD 209458\,b by \citet{snellen2010}, the cross-correlation technique has been commonly used to detect atomic and molecular species in the atmospheres of exoplanets from high-resolution spectra \citep[see][for an in-depth review of this method]{birkby2018}. Here, we followed a strategy similar to \citet{hoeijmakers2020wasp121b} to search for metals in the atmosphere of HAT-P-70\,b.

We first Doppler shifted the 2-dimensional spectra by the barycentric velocity of the Earth, consequently shifting the spectra from the rest frame of the observatory to that of the Solar System barycenter. We did not correct for the reflex motion of the star because its radial velocity semi-amplitude is only known to a $3\sigma$ upper limit of $K < 0.649 ~ \kms$. This upper limit is below the $\sim 0.8 ~ \kms$ that each pixel represents, and the star is rapidly rotating ($v \sin i_\star = 99.85^{+0.64}_{-0.61}~\kms$), which means that correcting for the stellar motion would not significantly impact the results \citep{casasayasbarris2018}. This choice also ensures the deep static \ion{Na}{1} features, most likely due to interstellar medium absorption, are removed in later stages of the analysis.

To keep the ratio between the average flux of the different orders constant, we normalized the average flux of each order to the average flux of that order over time. While this process removes information about the average flux in each exposure, this information is taken into account when combining the different in-transit exposures (see Section~\ref{sec:results_detections}). We then performed sigma-clipping to remove cosmics and other outliers that were not corrected in the data reduction process: after stacking all exposures, we ran a sliding 40-pixel window that replaced flux values farther than $5\sigma$ away from the mean by the interpolation of their neighboring pixels. Finally, we masked the spectral regions with evident sky emission and the spectral lines that were not accurately modeled by \texttt{molecfit}, notably the deep telluric lines in the oxygen B-band ($\sim 6900$~\AA\, see Figure~\ref{fig:molecfit}).

\subsubsection{Atmospheric Model Templates}\label{sec:templates}
The atmosphere of a planet filters the light of the star during the planet transit. The different gases that compose the atmosphere absorb the light at different wavelengths, causing the apparent size of the planet to change with wavelength. The transmission spectra models calculate how the planet size changes for various combinations of atmospheric gases and planet bulk parameters. In this work, we search for neutral and singly ionized metals with atomic number up to 28 and with significant spectral features in the wavelength range covered by HARPS-N. This includes \ion{Al}{1}, \ion{Ca}{1}, \ion{Ca}{2}, \ion{Co}{1}, \ion{Cr}{1}, \ion{Cr}{2}, \ion{Fe}{1}, \ion{Fe}{2}, \ion{K}{1}, \ion{Mg}{1}, \ion{Mn}{1}, \ion{Mn}{2}, \ion{Na}{1}, \ion{Ni}{1}, \ion{Sc}{1}, \ion{Sc}{2}, \ion{Si}{1}, \ion{Ti}{1}, \ion{Ti}{2}, \ion{V}{1} and \ion{V}{2}.  

We calculated the absorption cross-sections for the different gases using the open-source and custom opacity calculator \texttt{HELIOS-K} (\citealt{2021Grimm}). \texttt{HELIOS-K} is a very efficient opacity calculator that runs on graphics processing units (GPUs). As input to calculate the gas opacities, we used the line-list tables from \cite{2018Kurucz}, assumed Voigt line profiles for the absorption lines, 0.01 cm$^{-1}$ spectral resolution, and a constant line cutoff of 100 cm$^{-1}$. To calculate the transmission spectra, we developed our code based on the simple formalism presented in \cite{2017Gaidos} and \cite{2019Bower}. Our model computes the effective tangent height in an atmosphere that was discretized in 200 annuli. The chemical concentration calculations were done using the open-source code FastChem (\citealt{2018Stock}) assuming solar metallicities, except for the species \ion{Sc}{1} and \ion{Sc}{2} that are not included in the open-source version of FastChem. For \ion{Sc}{1} and \ion{Sc}{2}, we considered a constant mixing ratio of $10^{-8}$, which is a simple first-order approximation to search for possible detections. We included in our model the H$^{-}$ bound–free and free–free absorption from \cite{1988John}. Each high-resolution transmission spectrum includes one gas species along with H$^{-}$ continuum absorption and scattering by H and H$_2$. The surface gravity of the planet was set to $\log g = 3$, and the atmosphere was assumed to be in isothermal conditions, with a temperature of 4000~K. HAT-P-70\,b orbits an A-type star, and we therefore expect the atmospheric temperature at the pressures probed in transmission spectroscopy to be significantly higher than the planetary equilibrium temperature \citep{lothringer2019,stangret2020}.

\begin{figure*}
    \centering
    \includegraphics[width=\linewidth]{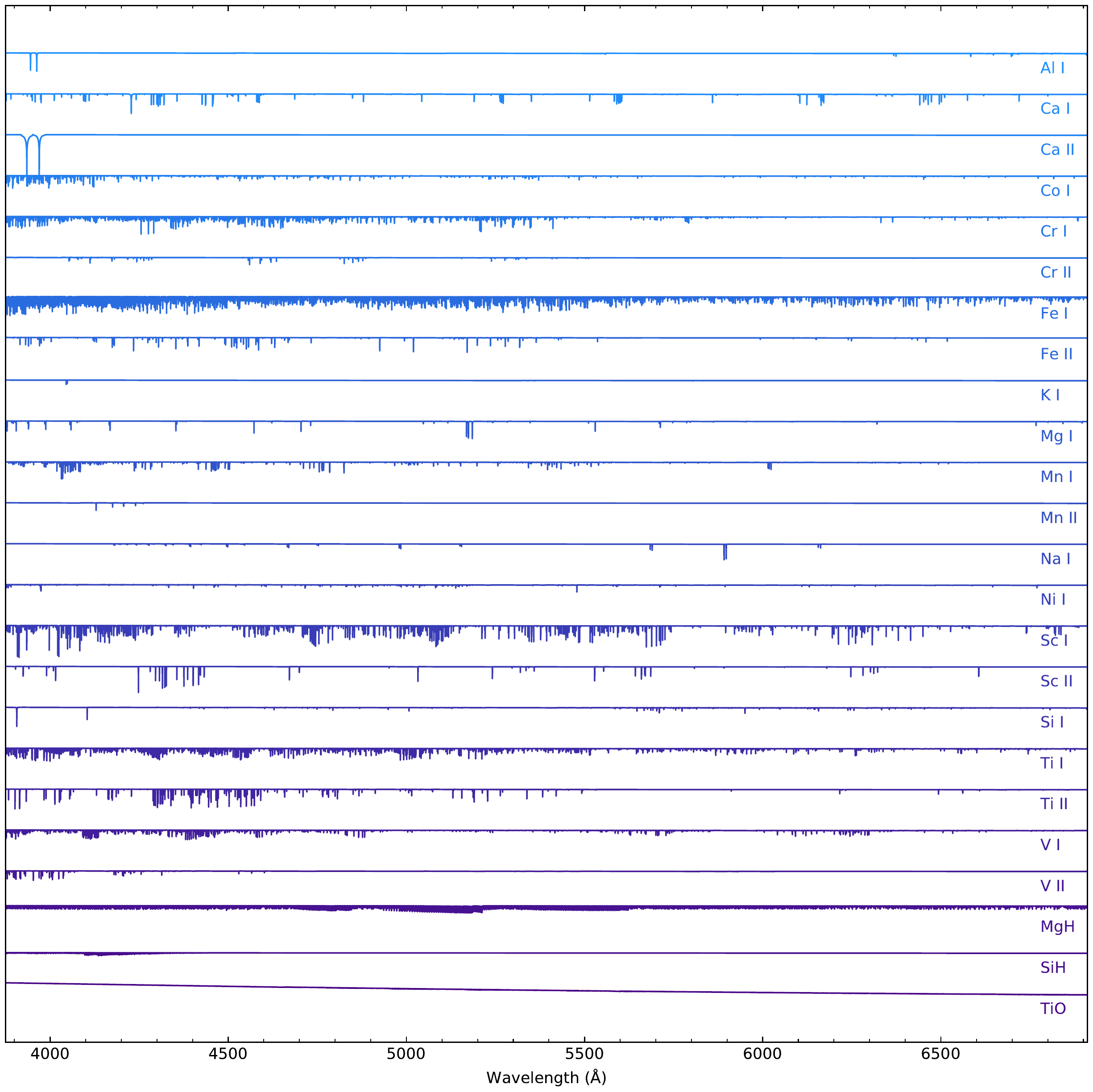}
    \caption{High-resolution models for each of the species surveyed in this work. These models are on the same scale (except for TiO, which is scaled up by a factor of 20), with a constant offset between each of them for clarity.}
    \label{fig:templates_all}
\end{figure*}

In addition to the atomic exploration, we also search for the presence of molecules using the ExoMol line-list database. We focus on MgH \citep{yadin2012}, SiH \citep{yurchenko2018} and TiO \citep{mckemmish2019}, as these species present multiple spectral features in the visible range and can potentially withstand the high temperatures of HAT-P-70\,b. We set the temperature of these models to 4000~K, except for MgH, which we set to 3000~K because there was no data available at 4000~K. We present these models, together with the atomic models, in Figure~\ref{fig:templates_all}.

The templates were then broadened using a Gaussian kernel to match the spectral resolving power of HARPS-N, $R=115,000$. The continuum, obtained by a quadratic spline interpolation of the highest values in each $30~$\AA\ window, is then removed from each model template. Those values smaller than $1\times10^{-4}$ times the value of the deepest line in each template were set to zero \citep{hoeijmakers2020wasp121b}. Following subtraction of the continuum, these models served as cross-correlation templates.

\subsubsection{Cross-Correlations with the Model Templates}\label{sec:cc_method}
The cross-correlation function coefficients $c(v,t)$ were calculated according to:
\begin{equation}\label{eq:correlations}
c(v,t) = {\sum_{i=0}^{N}{x_i(t) T_i(v)}},
\end{equation}
where $x_i(t)$ is the flux at the i-th pixel in the spectrum obtained at time $t$, and $T_i(v)$ is the atmospheric template Doppler-shifted by a velocity $v$, which takes values ranging from $-500$ to $500~\kms$ in steps of $1~\kms$. $N$ indicates the total number of pixels in each spectrum, accross all orders. The spectra were neither continuum normalized nor blaze corrected. Assuming photon noise is the dominant noise source, the weight of each pixel in the sum of Equation~\ref{eq:correlations} is therefore proportional to its variance, without the need for explicit weights \citep{hoeijmakers2020wasp121b}. To obtain the cross-correlation functions (CCFs) corresponding to the transmission signal, all coefficients were subsequently divided by the average of those located out of transit. We applied a Gaussian filter of width 70 \kms\ to remove any remaining broadband variations \citep{hoeijmakers2020wasp121b}.

In the resulting CCFs, the radial velocity of the planet $v_{p} \left(t\right)$ is given by
\begin{equation}\label{eq:vplanet}
    v_p \left(t\right) = K_p \sin{\left(2\pi\phi(t)\right) + v_{\textup{sys}}},
\end{equation}
where $K_{p}$ is the planet radial velocity semi-amplitude, $\phi(t)$ is the orbital phase at time $t$, and $v_{\textup{sys}}$ is the systemic velocity.

\subsubsection{Spin-Orbit Alignment}\label{sec:dopshad_method}
During the planetary transit, the rotationally broadened line profiles of the stellar spectrum are distorted by the planet blocking part of the stellar disk. This generates a time dependent signal in the CCFs of those species present in the star, known as the Doppler shadow \citep{colliercameron2010}. The Doppler shadow could potentially bias the results from the cross-correlation analysis, and therefore we removed it from the CCFs --- for this purpose, we performed a direct spectral modeling approach, which we describe in Section~\ref{sec:rm_model}. Here, we focus on how we used the Doppler shadow as a powerful source of information about the architecture of the system. In particular, we used the location of the Doppler shadow in each CCF to derive the angle between the orbital axis of the planet and the spin axis of the star, also known as spin-orbit angle or obliquity.

For each epoch, the velocity in the cross-correlation function at which the Doppler shadow appears corresponds to the line-of-sight velocity of the part of the stellar disk occulted by the planet. Ignoring differential rotation and convection, this velocity is given by
\begin{equation}
    v_{\textup{occ}}(t) = x_\perp v \sin{i_\star},
\end{equation}
where $v \sin{i_\star}$ is the apparent rotational velocity of the star, and $x_\perp$ is the orthogonal distance from the planet to the stellar spin axis. This distance can be expressed as
\begin{equation}\label{eq:x_orth}
    x_\perp = \frac{a}{R_\star}\left(\sin{(2\pi\phi)}\cos{\lambda} +
    \cos{(2\pi\phi)}\cos{i_p}\sin{\lambda}\right),
\end{equation}
where $a$ is the orbital semi-major axis, $R_\star$ is the stellar radius, $\lambda$ is the obliquity and $i_p$ is the orbital inclination \citep{cegla2016}.

\begin{figure}
    \centering
    \includegraphics[width=\linewidth]{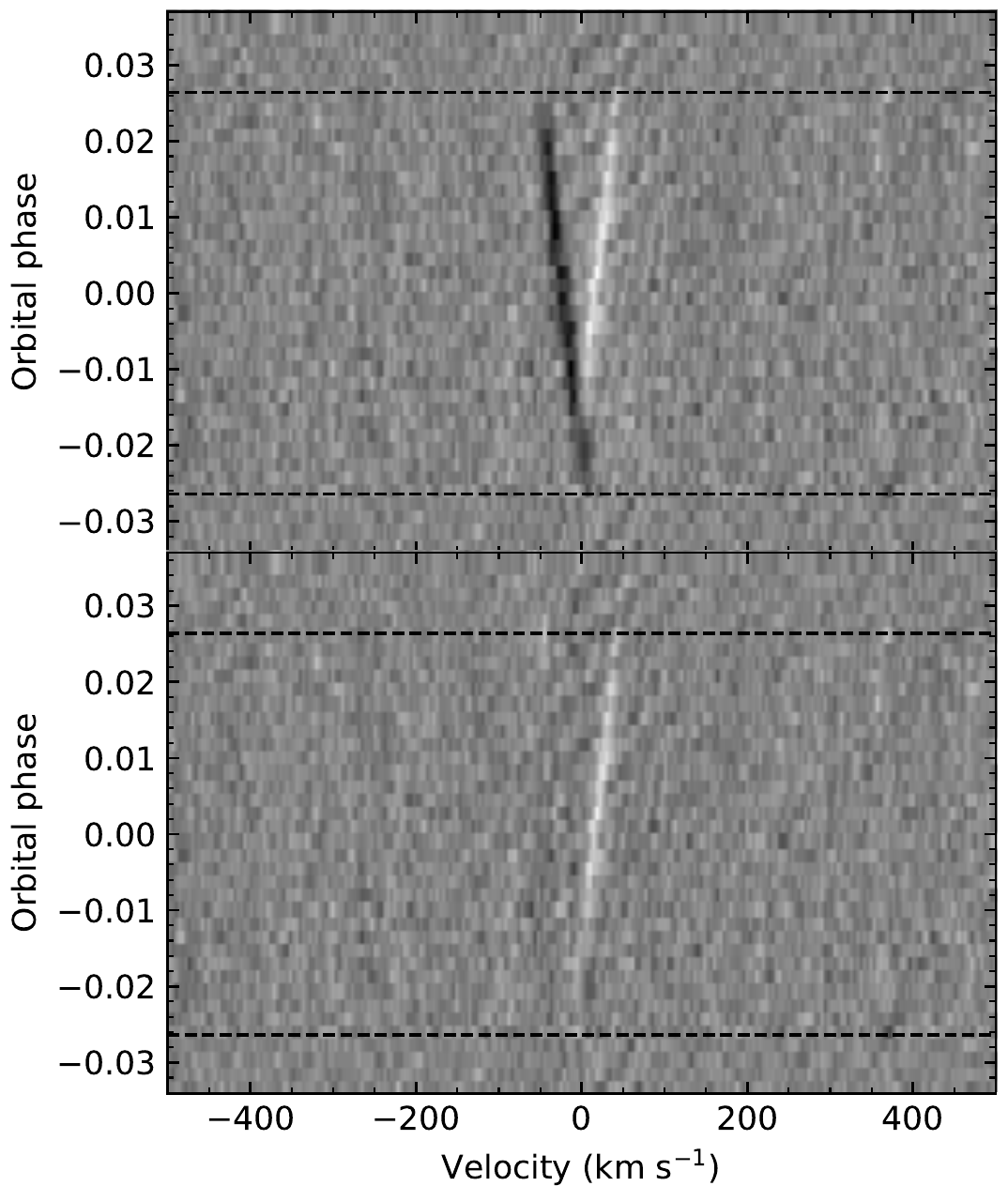}
    \caption{Removal of the Doppler shadow. \textit{Top:} CCFs that result from cross-correlating the different exposures with a PHOENIX template and dividing by the master out-of-transit CCF. The black dashed lines mark the beginning and end of the transit. The black slanted streak, visible during transit, is caused by the planet blocking part of the stellar disk and is known as the Doppler shadow. The white slanted streak is the planet atmosphere signal. \textit{Bottom:} Same as above, but in this case the spectra were corrected for the Rossiter-McLaughlin and Center-to-Limb Variation effects priors to cross-correlation, as described in Section~\ref{sec:rm_model}, which caused the Doppler shadow to disappear.}
    \label{fig:dopshad}
\end{figure}

Following \citet{hoeijmakers2019}, we used a high-resolution synthetic spectrum from the PHOENIX library \citep{husser2013} with an effective temperature of $T_{\textup{eff}} = 8400~\textrm{K}$ and surface gravity of $\log g = 4.00$, values that closely resemble those of the host star. This PHOENIX template was then treated similarly to the atmospheric model templates: it was continuum normalized and broadened to match the instrumental resolution of HARPS-N, and it was used thereafter to produce the CCFs. Cross-correlating the spectra with a PHOENIX template yields two strong signals: the Doppler shadow and the exoplanet atmosphere signal, as shown in Figure~\ref{fig:dopshad}.
We approximated the Doppler shadow in each in-transit CCF with a Gaussian function, ignoring the exposures where its velocity was close to that of the atmosphere signal.

We then conducted a Markov Chain Monte Carlo (MCMC) analysis of the Doppler shadow, similar to that in \citet{hoeijmakers2020expres}. We used the centroids $v_{occ}^{\textup{obs}}(t)$ and centroid uncertainties $\sigma_{v}^{obs}(t)$ from the Gaussian fit of the Doppler shadow in each in-transit CCF, and assumed a Gaussian log-likelihood function
\begin{equation}\label{eq:logl}
    L \propto -\frac{1}{2}\sum_t\left(\frac{v_{\textup{occ}}^{\textup{obs}}(t)-v_{\textup{occ}}^{\textup{mod}}(t)}{\sigma_{v}^{\textup{obs}}(t)}\right)^2,
\end{equation}
where $v_{\textup{occ}}^{\textup{mod}}(t)$ is the line-of-sight velocity of the part of the stellar disk occulted by the planet at time $t$. Our model parameters include $v\sin{i_\star}$, $a/R_\star$, $\lambda$, $i_p$ and $v_{\rm{sys}}$. We used a uniform prior for the obliquity ($0^\circ < \lambda < 360^\circ$) and Gaussian priors for the parameters derived from the transit light curve analysis in \citet{zhou2019} ($a/R_\star$ and $i_p$), with the mean and standard deviation shown in Table~\ref{tab:mcmc}. We also used Gaussian priors for $v\sin{i_\star}$ and for $v_{\rm{sys}}$, each with a conservative standard deviation of 5~\kms\ to account for potential offsets from the spectroscopic measurements in \citet{zhou2019}. We sampled the posterior distributions with the \texttt{emcee} Python package \citep{emcee}.

\subsubsection{Modeling the Rossiter-McLaughlin and Center-to-Limb Variation Effects}\label{sec:rm_model}
The spectrum across a stellar disk is not uniform. The spectrum originating in each region of the stellar disk is Doppler-shifted by an amount proportional to the projected rotational velocity and the distance to the stellar spin axis. Additionally, the line profile changes as a function of the limb angle $\mu = \cos{\theta}$, where $\theta$ is the angle between the line of sight and the normal to the stellar surface. Recent works in exoplanet transmission spectroscopy have highlighted the importance of correcting for these two effects, known as the Rossiter-McLaughlin and Center-to-Limb Variation effects, as they can significantly impact the results, potentially leading to false negatives and false positives \citep[e.g.][]{czesla2015,yan2017,chen2020,casasayasbarris2021}.

When combining the different transmission spectra in the rest frame of a planet on a near-polar orbit such as HAT-P-70\,b, the spectral signature produced by the Rossiter-McLaughlin effect largely smears out. The Center-to-Limb Variation effect is also unlikely to substantially alter the results of our analysis, as this effect is generally weaker in hot stars and HAT-P-70 has one of the highest effective temperatures among all known exoplanet hosts. For completeness, we still chose to model these effects, following the steps outlined in \citet{cabot2020}.

Using \texttt{Spectroscopy Made Easy} version 522 (SME, \citealt{piskunov2017}), the Vienna Atomic Line Database (VALD, \citealt{ryabchikova2015}) and the Kurucz \texttt{ATLAS9} model atmospheres \citep{castelli2003}, we computed stellar spectra at 21 limb angles, ranging from the center of the stellar disk to the limb in steps of $\Delta\mu = 0.05$. We used the stellar parameters in \citet{zhou2019}. We divided the stellar disk in square cells with side length $0.01~R_\star$ and assigned to each cell a spectrum that we obtained from linear interpolation of the 21 reference spectra. The spectrum of each cell is Doppler-shifted by the line-of-sight velocity of that cell assuming no differential rotation. 

At the orbital phase of each exposure we integrated the spectra in the cells blocked by the planet disk. The radius of this disk was set to that of HAT-P-70\,b and its coordinates were calculated with Equation~\ref{eq:x_orth} and 
\begin{equation}
    y_\perp = \frac{a}{R_\star}\left(\sin{(2\pi\phi)}\sin{\lambda} -
    \cos{(2\pi\phi)}\cos{i_p}\cos{\lambda}\right),
\end{equation}
using the parameters derived from the Doppler shadow analysis in Section~\ref{sec:dopshad_method}. These integrated spectra were then divided by the stellar spectrum (obtained from the integration of the spectra in all cells of the stellar disk), continuum normalized using a cubic polynomial and broadened to match the resolution of HARPS-N. We then used these spectra to correct the 2-dimensional spectra prior to cross-correlation, which resulted in the removal of the Doppler shadow from the CCFs (see Figure~\ref{fig:dopshad}).

\subsection{Transmission Spectroscopy}
Often, the discoveries of atomic species in ground-based, high-resolution, exoplanet transmission spectra involve a cross-correlation approach to combine the S/N of multiple lines. However, some species can be identified through a single prominent spectral feature, such as the \ion{Na}{1} doublet, the \ion{Ca}{2} H \& K lines, and the \ion{H}{1} Balmer lines. To search for these prominent spectral lines, we derived the transmission spectrum of HAT-P-70\,b. Our analysis was similar to that of \citet{cabot2020}, with slight modifications as described below.

After performing sigma-clipping on the 2-dimensional spectra, we divided each order by the corresponding blaze function and Doppler-shifted them to the HAT-P-70 rest frame using the systemic velocity and the Earth's barycentric velocity. We did not correct for the reflex motion of the star due to the reasons outlined in Section~\ref{sec:metalsearch}. As in the cross-correlation analysis, we corrected for the Rossiter-McLaughlin and Center-to-Limb Variation effects using the spectra computed in Section~\ref{sec:rm_model}.

We produced the master stellar spectrum by combining all out-of-transit spectra:
\begin{equation}
    \hat{f}_{\textup{out}}(\lambda) = \sum_{t_\textup{out}}
    w(\lambda,t_\textup{out}) f(\lambda,t_\textup{out}),
\end{equation}
where $f(\lambda,t_\textup{out})$ corresponds to the flux at the wavelength $\lambda$ at an out-of-transit time $t_\textup{out}$. We weighted these fluxes by $w(\lambda,t_\textup{out})$, the inverse of their squared uncertainties (calculated assuming photon noise) to account for the changes in S/N throughout the observations. 

Each spectrum was then divided by this master out-of-transit spectrum to produce the transmission spectra:
\begin{equation}
    \mathfrak{R}(\lambda, t) = \frac{f(\lambda,t)}{\hat{f}_{\rm out}(\lambda)}.
\end{equation}
To normalize the transmission spectra, we fit a linear function to each spectrum and divided it out. \citet{cabot2020} used a 5$^{\rm th}$ degree polynomial in the normalization process, but their analysis involved the stitched spectrum that combines all orders from HARPS. Our analysis is on an order-by-order basis, so a linear polynomial suffices. We found significant residuals in the core of the stellar \ion{Na}{1} doublet. We masked out these residuals during the analysis to not influence the results.

We then Doppler-shifted all transmission spectra to the planet rest-frame using the planet radial velocity given by Equation~\ref{eq:vplanet}. We obtained the master transmission spectrum by a weighted average of the in-transit normalized transmission spectra. We removed any residual broadband variations using a median filter with a width of 1501 pixels \citep{cabot2020}.

\section{Results and discussion} \label{sec:resultsdiscussion}

\subsection{A Planet on a Nearly Polar Orbit}
\begin{figure}
    \centering
    \includegraphics[width=\linewidth]{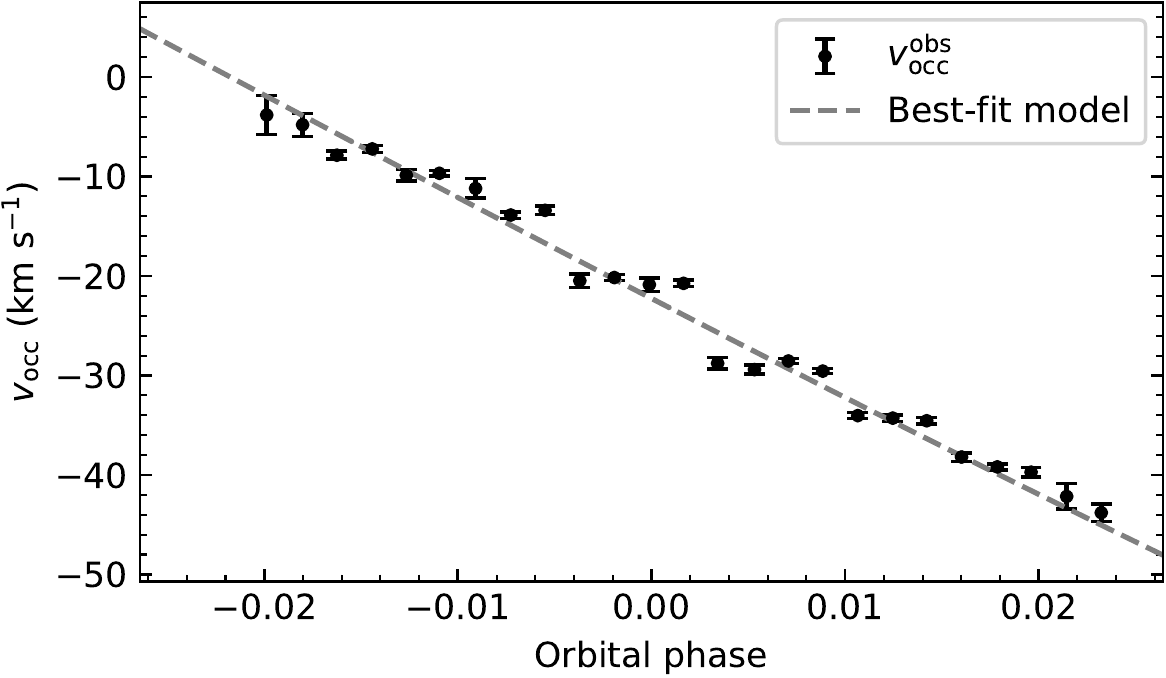}
    \caption{Best-fit model of the Doppler shadow. The black dots correspond to the centroid velocities of the Gaussian fits to each of the CCFs that make up the Doppler shadow of Fig.~\ref{fig:dopshad}. To avoid contamination from the atmosphere of HAT-P-70\,b, we discard measurements at velocities that are close to the radial velocity of the exoplanet signal. The grey dashed line marks the best-fit model that results from the evaluation of Eq~\ref{eq:logl}.}
    \label{fig:bfm}
\end{figure}

Our MCMC analysis of the Doppler shadow, described in Section \ref{sec:dopshad_method}, consists of 100 walkers of 100,000 steps each. We discard the first 5,000 steps to avoid sampling during the burn-in phase, and thin the chains by 800 (approximately the auto-correlation time). Table~\ref{tab:mcmc} lists the values derived from our analysis, and Figure~\ref{fig:bfm} presents the measurements of the centroid velocities and the best-fit model. Figure~\ref{fig:posteriors} shows the posterior distributions. 

\begin{table*}
\centering
\begin{tabular}{l|ccrr}
\hhline{=====}
Parameter & Symbol & Unit & Prior & Posterior \\
\hline
Scaled semi-major axis & $a/R_\star$ & – & $\mathcal{N}(5.45,0.49)$ & $5.23^{+0.46}_{-0.47}$ \\
Orbital inclination & $i_p$ & $^\circ$ & $\mathcal{N}(96.5,1.4)$ & $95.47^{+0.80}_{-0.73}$  \\
Obliquity & $\lambda$ & $^\circ$ & $\mathcal{U}(0, 360)$ & \obliq \\
Projected stellar rotational velocity & $v\sin{i_\star}$ & $\rm km~s^{-1}$  & $\mathcal{N}(99.9,5.0)$ & $98.6^{+5.0}_{-4.9}$  \\
Systemic velocity & $v_{\rm{sys}}$ & $\rm km~s^{-1}$  & $\mathcal{N}(25.2,5.0)$ & $25.0^{+4.6}_{-4.7}$ \\
\hline
\end{tabular}
\caption{Results of the MCMC analysis of the Doppler shadow. $\mathcal{N}(\mu,\sigma)$ indicates a Gaussian distribution with mean $\mu$ (from \citealt{zhou2019}) and standard deviation $\sigma$, while $\mathcal{U}(a,b)$ denotes a uniform distribution from $a$ to $b$. The last column shows the median value derived for each parameter. The 16th and 84th percentiles are used to establish the uncertainties. }\label{tab:mcmc}
\end{table*}

\citet{zhou2019} find an obliquity of $\lambda = {113.1^\circ}^{+5.1^\circ}_{-3.4^\circ}$ by simultaneously modeling the light curve and the Doppler shadow. According to the classification in \citet{addison2018}, this value corresponds to a retrograde orbit ($112.5^\circ \leq \left\lvert\lambda\right\lvert \leq 247.5^\circ$). In this work, we update the value of the obliquity of HAT-P-70\,b to $\lambda = \obliqwithunits$, indicative of a nearly polar orbit ($67.5^\circ < \left\lvert\lambda\right\lvert < 112.5^\circ$ and $247.5^\circ < \left\lvert\lambda\right\lvert < 292.5^\circ$). Like many other hot Jupiters found around hot stars, HAT-P-70\,b shows a highly misaligned orbit \citep{winn2010}.

\subsection{Results from the Cross-Correlation Analysis}
From our cross-correlation analysis, we claim the detection (i.e. a signal with $\rm{S/N}>4$ at a location consistent with that of the planet) of the following neutral and ionized species in the atmosphere of HAT-P-70\,b: \ion{Ca}{2}, \ion{Cr}{1}, \ion{Cr}{2}, \ion{Fe}{1}, \ion{Fe}{2}, \ion{Mg}{1}, \ion{Na}{1} and \ion{V}{1}. We also present tentative evidence (i.e. with $3 \leq \rm{S/N} \leq 4$) of \ion{Ca}{1} and \ion{Ti}{2} that will require further investigation.

\subsubsection{Detection of \ion{Ca}{2}, \ion{Cr}{1}, \ion{Cr}{2}, \ion{Fe}{1}, \ion{Fe}{2}, \ion{Mg}{1}, \ion{Na}{1} and \ion{V}{1}}\label{sec:results_detections}

\begin{figure*}
    \centering
    \includegraphics[width=\linewidth]{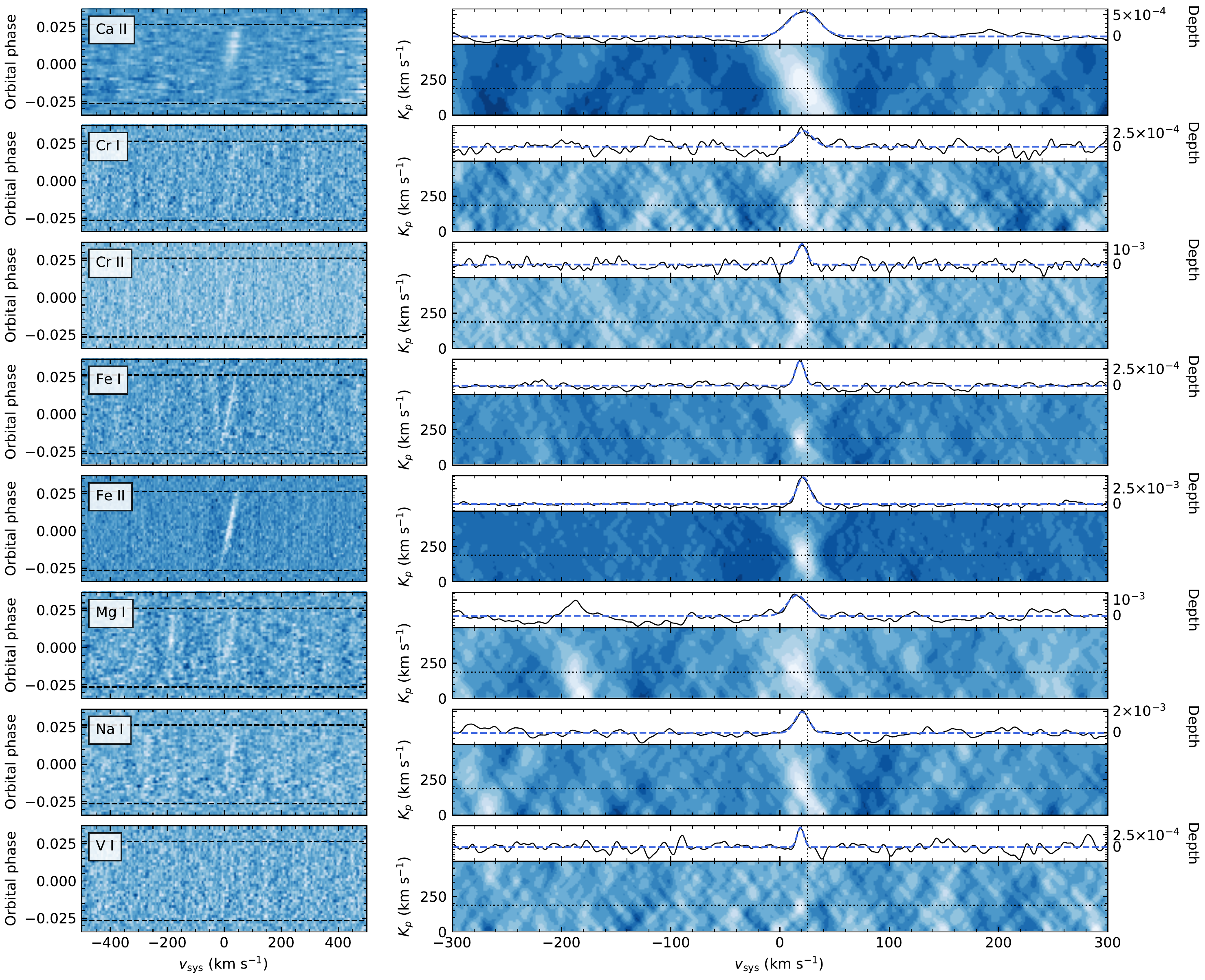}
    \caption{Detection of various neutral and ionized species in the atmosphere of HAT-P-70\,b. \textit{Left:} CCFs of the species for which we claim a detection (\ion{Ca}{2}, \ion{Cr}{1}, \ion{Cr}{2}, \ion{Fe}{1}, \ion{Fe}{2}, \ion{Mg}{1}, \ion{Na}{1} and \ion{V}{1}). The CCFs are in the rest frame of the Solar System barycenter. The Doppler shadow has already been removed. The horizontal dashed lines mark the beginning and end of the transit. The slanted white streak visible during the transit corresponds to the planet atmosphere. \textit{Right:} $K_p-v_{\rm sys}$ maps, with the in-transit CCFs co-added at different planet radial velocity semi-amplitudes (i.e. along different slopes) and at different systemic velocities. The dotted lines indicate the expected systemic velocity and radial velocity semi-amplitude of the planet \citep{zhou2019}.  The values at the expected planet radial velocity semi-amplitude, $K_p = 187 ~ \kms$, are plotted above each $K_p-v_{\rm sys}$ map, fitted with a Gaussian model. In the case of \ion{Mg}{1}, the peak near $v_{\rm sys}=-190~\kms$ is caused by contamination from one of the strongest \ion{Fe}{2} lines.}
    \label{fig:ccfs}
\end{figure*}

In this section, we present the detection of \ion{Ca}{2}, \ion{Cr}{1}, \ion{Cr}{2}, \ion{Fe}{1}, \ion{Fe}{2}, \ion{Mg}{1}, \ion{Na}{1} and \ion{V}{1} in HAT-P-70\,b via the cross-correlation method. The panels in the left column of Figure~\ref{fig:ccfs}
show the CCFs that result from cross-correlating the HARPS-N data set with model templates of these species. The slanted white streak, discernible during transit, corresponds to the atmospheric signal.

To combine the signal from the in-transit CCFs, we construct $K_{p}-v_{\rm sys}$ diagrams by taking the weighted average of the in-transit cross-correlation coefficients along different planet radial velocity semi-amplitudes $K_p$ and systemic velocities $v_{\rm sys}$ \citep{brogi2012}. The weights correspond to the mean flux in each exposure \citep{hoeijmakers2020wasp121b}. We sample planet radial velocity semi-amplitudes between 0 and $500~\kms$ and systemic velocities between -300 and $300~\kms$, in steps of $1~\kms$. 

The resulting $K_{p}-v_{\rm sys}$ diagrams are presented in the right column of Figure~\ref{fig:ccfs}. We convert the cross-correlation coefficients to S/N values by dividing them by the standard deviation of the coefficients located at systemic velocities farther than $50~\kms$ from where we expect the planet signal \citep[e.g.][]{brogiline2019,stangret2020}. We measure enhancements in the cross-correlation coefficients near the predicted location of the exoplanet, with S/N values ranging from 4.7 (\ion{V}{1}) to 17.6 (\ion{Fe}{2}). In Table~\ref{tab:ccf_fits},
\begin{table*}
\centering
\begin{tabular}{l||rr|rrrr}
\hhline{=======}
 & \multicolumn{2}{c|}{Strongest signal} & \multicolumn{4}{c}{Gaussian fit at $K_p=187~\kms$} \\
\hhline{~------}
Species & S/N$_{\rm{max}}$ & \{$K_p$, $v_{\rm{sys}}$\}$_{\rm{max}}$ & A ($\times 10^{-3}$) & $\Delta v_{\rm{sys}}$ $(\rm km~s^{-1})$ & FWHM $(\rm km~s^{-1})$ & $\Delta R_p / R_p$ \\
\hline
\ion{Ca}{1} $^{*}$ & 3.2 & $\{144^{+65}_{-44}, 21^{+5}_{-3}\}$   & $0.39 \pm 0.12$ & $-3.0 \pm 1.2$ & $8.2 \pm 2.9$ & $0.020\pm0.006$ \\
\ion{Ca}{2} & 9.0 & $\{236^{+74}_{-163}, 18^{+15}_{-5}\}$ & $0.59 \pm 0.03$ & $-3.8 \pm 0.8$ & $31.5 \pm 1.9$ & $0.030\pm0.002$\\
\ion{Cr}{1} & 4.8 & $\{172^{+59}_{-90}, 20^{+5}_{-3}\}$ & $0.28 \pm 0.05$ & $-3.4 \pm 1.7$ & $19.3 \pm 4.0$ & $0.014\pm0.003$\\
\ion{Cr}{2} & 6.6 & $\{154^{+27}_{-17}, 20^{+1}_{-2}\}$ & $1.40 \pm 0.22$ & $-5.2 \pm 0.8$ & $10.4 \pm 1.9$ & $0.069\pm0.011$\\
\ion{Fe}{1} & 11.1 & $\{169^{+21}_{-18}, 19^{+1}_{-2}\}$  & $0.37 \pm 0.03$ & $-7.1 \pm 0.4$ & $9.7 \pm 1.0$ & $0.019\pm0.002$\\
\ion{Fe}{2} & 17.6 & $\{177^{+34}_{-24}, 21^{+1}_{-2}\}$ & $4.37 \pm 0.17$ & $-3.9 \pm 0.3$ & $13.7 \pm 0.6$ & $0.203\pm0.009$ \\
\ion{Mg}{1} & 5.2 & $\{192^{+47}_{-91}, 13^{+11}_{-4}\}$ & $1.31 \pm 0.18$ & $-9.6 \pm 1.4$ & $20.7 \pm 3.3$ & $0.065\pm0.009$\\
\ion{Na}{1} & 6.1 & $\{180^{+61}_{-53}, 21^{+2}_{-4}\}$ & $1.95 \pm 0.24$ & $-5.2 \pm 0.9$ & $15.1 \pm 2.1$ & $0.095\pm0.011$\\
\ion{Ti}{2} $^{*}$ & 3.6 & $\{102^{+201}_{-45}, 13^{+5}_{-5}\}$ & $0.49 \pm 0.10$ & $-13.2 \pm 1.8$ & $17.4 \pm 4.2$ & $0.025\pm0.005$ \\
\ion{V}{1} & 4.7 & $\{161^{+44}_{-26}, 19^{+2}_{-2}\}$ & $0.40 \pm 0.09$ & $-6.5 \pm 0.8$ & $7.3 \pm 2.0$ & $0.020\pm0.005$\\
\hline
\end{tabular}
\caption{Results from the cross-correlation analysis for the species for which we claim a detection and for the species for which we only find tentative evidence (marked with an asterisk). For each species we indicate the maximum S/N in the $K_p-v_{\rm{sys}}$ diagram (S/N$_{\rm{max}}$) and the velocity pair (\{$K_p$, $v_{\rm{sys}}$\}$_{\rm{max}}$) at which this maximum S/N occurs. We also indicate the best-fit Gaussian parameters (amplitude, velocity offset and full width at half maximum) of the CCFs at the expected planet radial velocity semi-amplitude, $K_p = 187 ~ \kms$ \citep{zhou2019}, and express the amplitude as an excess planet radius in units of planet radii ($\Delta R_p / R_p$).}\label{tab:ccf_fits}
\end{table*}
we present the velocity pair (i.e. $K_p$ and $v_{\rm sys}$) and the peak S/N values for each of the species for which we claim a detection.

The co-added cross-correlation coefficients at the expected radial velocity semi-amplitude of the exoplanet \citep[$K_p = 187~\kms$, calculated according to the orbital parameters in][]{zhou2019} are plotted above each $K_{p}-v_{\rm sys}$ diagram. We fit these cross-correlation coefficients by a Gaussian function using the Levenberg-Marquardt algorithm to determine the depth, Doppler shift and width of each signal. As in \citet{hoeijmakers2019}, we set the uncertainty in the data as the standard deviation of the values at $\left|\Delta v_{\rm sys}\right| > 50~\kms$, far from the planetary signal. Our Gaussian function fit only includes every $4^{\rm th}$ data point. This is because the atmospheric model templates were broadened to match the spectral resolution of HARPS-N (see Section~\ref{sec:templates}), meaning that the correlation length in the resulting CCFs is $\Delta v_{\rm{sys}} \approx 3~\kms$ \citep{colliercameron2010, hoeijmakers2020wasp121b}. Table~\ref{tab:ccf_fits} summarizes the results of the Gaussian fits.

The amplitudes of the different signals span an order of magnitude, from $0.28\times10^{-3}$ (\ion{Cr}{1}) to $4.37\times10^{-3}$ (\ion{Fe}{2}). These amplitudes indicate the excess transit depth associated with each species. We can express this excess transit depth, $\Delta \delta$, as an excess planet radius, $\Delta R_p$:
\begin{equation}\label{eq:depth}
    \frac{\Delta R_p}{R_p}=\sqrt{\frac{(R_p/R_\star)^2+\Delta\delta}{(R_p/R_\star)^2}}-1.
\end{equation}
Table~\ref{tab:ccf_fits} shows the excess planet radius of each species. These measurements provide an estimate of the radius at which the cores of a weighted average of the spectral lines become optically thick, and therefore depend on the line opacities and the abundance profile of each species.

Various factors contribute to the broadening of the lines in the transmission spectrum of exoplanet atmospheres, including winds, instrumental broadening, Doppler broadening caused by the atomic thermal motion, and pressure broadening due to collisions with other particles. The rotation of the planet can also broaden the line profiles to a level detectable by high-resolution spectroscopy \citep{snellen2014}. Given the tidal forces expected at such short orbital distances, HAT-P-70\,b is likely to have a synchronous rotation such that its orbital period and rotational period are the same \citep{rasio1996,brogi2016}. This would induce a rotational broadening of $\sim 3.5~\rm{km~s^{-1}}$. Additionally, according to Equation~\ref{eq:vplanet}, the planet changes radial velocity by about $\sim 2~\rm{km~s^{-1}}$ during each exposure, which would further contribute to smearing out the signal. Thanks to the remarkable strength of the \ion{Fe}{1} and \ion{Fe}{2} signals in HAT-P-70\,b, we are able to measure their FWHMs with high precision. We find that the signal of \ion{Fe}{2} is significantly broader than that of \ion{Fe}{1}, potentially caused by the larger velocity range at the low pressures probed by \ion{Fe}{2} \citep{gibson2020}.

The centroid measurements in Table~\ref{tab:ccf_fits} indicate that all signals are blue-shifted from the exoplanet rest-frame by a few \kms. This blue-shift might be due to the presence of strong winds flowing at low pressures from the day-side to the night-side of the planet, driven by the large day-night heating gradients in extremely irradiated gas giants like HAT-P-70\,b \citep{showman2013,komacek2016}. Additionally, the blue-shifts are slightly different for the different species. For example, there is a significant difference in the blue-shifts of the \ion{Fe}{1} and \ion{Fe}{2} signals, with \ion{Fe}{1} experiencing a blue-shift $\sim3\,\kms$ larger than that of \ion{Fe}{2}. As the signals originate at different altitudes in the atmosphere, these differences can be explained by a stratification of the atmosphere \citep{nugroho2020,hoeijmakers2020expres}. Blue-shifts larger in \ion{Fe}{1} than in \ion{Fe}{2} have also been reported in the transmission spectrum of other ultra-hot Jupiters, including MASCARA-2\,b \citep{casasayasbarris2019,stangret2020,nugroho2020,hoeijmakers2020expres} and WASP-121\,b \citep{bourrier2020,gibson2020,ben-yami2020,hoeijmakers2020wasp121b,borsa2021}. A significantly blue-shifted spectral signature of \ion{Fe}{1} has also been detected in the ultra-hot Jupiter WASP-76\,b \citep{ehrenreich2020,tabernero2020}. Per contra, the \ion{Fe}{1} and \ion{Fe}{2} detections in KELT-9\,b show no significant Doppler-shift with respect to the rest frame of the planet \citep{hoeijmakers2019}. \citet{cabot2021} recently reported a detection of \ion{Fe}{1} in the atmosphere of the ultra-hot Jupiter TOI-1518\,b, blue-shifted by $\sim 2~\kms$, and a tentative detection of \ion{Fe}{2}, with a blue-shift of $\sim 4~\kms$.

We also note that determining the Doppler-shifts of the atmospheric spectral signatures with respect to the planet relies on the accurate measurement of the systemic radial velocity of the star. The systemic velocity used in our analysis is from \citet{zhou2019}, derived by modeling the stellar line profiles from a least-squares deconvolution using spectra taken by the Tillinghast Reflector Echelle Spectrograph (TRES). As pointed out by \citet{hoeijmakers2019}, different radial velocity derivation methods for rapidly rotating stars could result in differences in the systemic velocity of a few $\kms$. Our Doppler shadow analysis provides an alternative, less precise value for the systemic velocity of the system. This value is still larger than the systemic velocities at which we detect the different species, giving us confidence that the blue-shifts are likely not due to the determination of the systemic velocity.

\subsubsection{Tentative evidence of \ion{Ca}{1} and \ion{Ti}{2}}
\begin{figure*}
    \centering
    \includegraphics[width=\linewidth]{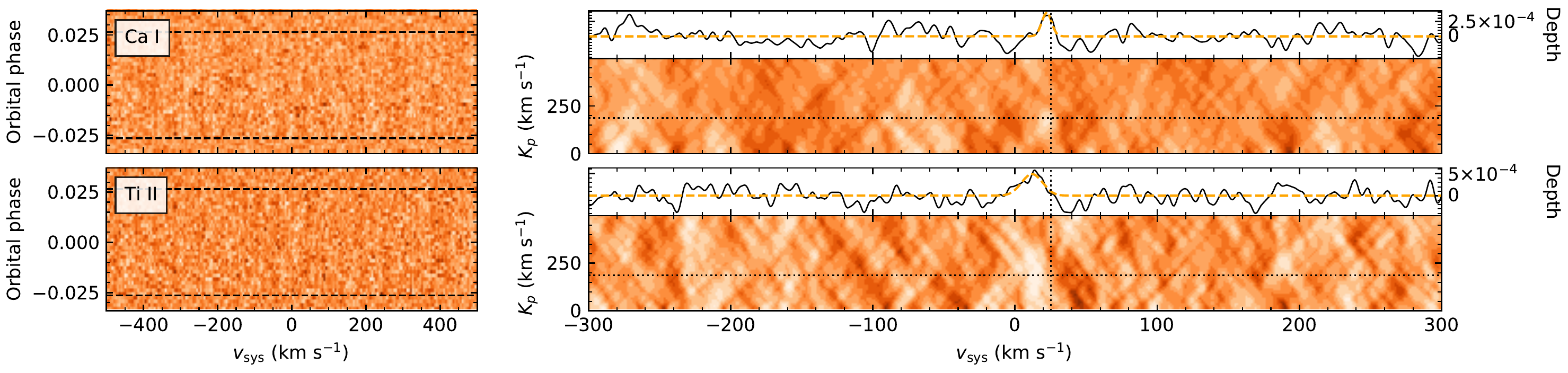}
    \caption{Same as Figure~\ref{fig:ccfs}, but for the species for which we find tentative evidence rather than claiming a detection. }
    \label{fig:ccfs_evidence}
\end{figure*}

Besides the detections presented above, we also find tentative evidence of \ion{Ca}{1} and \ion{Ti}{2} (Figure~\ref{fig:ccfs_evidence}).
The velocity pair of the \ion{Ca}{1} signal is consistent with that of the planet and of the other detected species. However, its S/N peaks at 3.2, which is below our detection threshold of 4. Also below the detection threshold is the signal of \ion{Ti}{2}, which peaks at S/N$_{\rm{max}}=3.6$. The \ion{Ti}{2} signal has a $K_p$ value consistent with that of the planet, and it is markedly extended and blue-shifted in the $K_p-v_{\rm{sys}}$ map. The CCFs of \ion{Ti}{2} present significant noise near $v_{\rm{sys}}=0~\kms$, which affects the $K_p-v_{\rm{sys}}$ map.

\subsection{Results from the Transmission Spectroscopy Analysis}

\subsubsection{Detection of \ion{Ca}{2} H \& K Lines}\label{sec:ca_lines}
\begin{figure*}
    \centering
    \includegraphics[width=\linewidth]{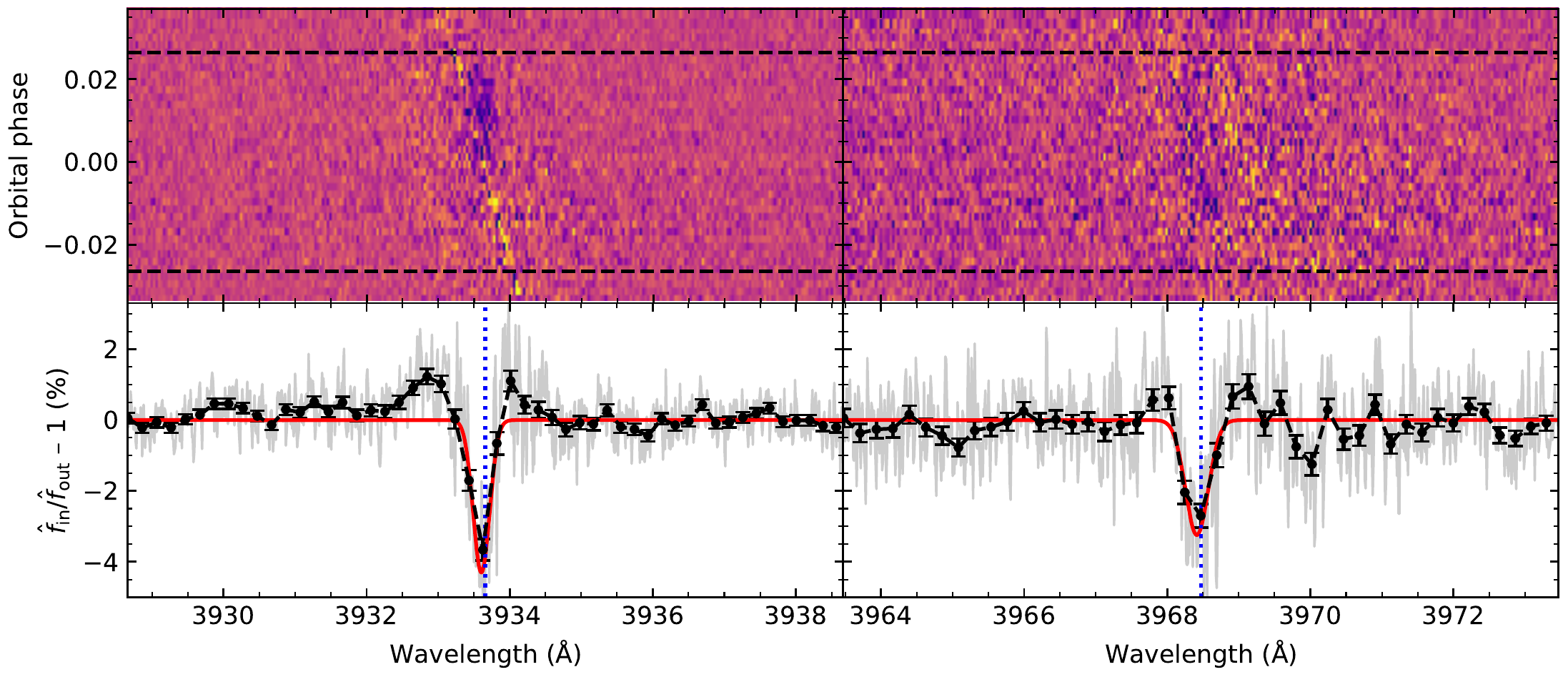}
    \caption{Transmission spectroscopy for the \ion{Ca}{2} K (left) and H (right) lines. \textit{Top row:} Residuals in the exoplanet rest-frame, obtained by dividing the spectrum at each orbital phase by the master out-of-transit spectrum and correcting for the contribution from the RM and CLV effects. The horizontal dashed lines mark the beginning and end of the transit. \textit{Bottom row:} Unbinned (gray) and 20x binned (black) master transmission spectrum, fitted with a Gaussian model (red). The vertical blue dotted line indicates the theoretical location of each spectral line. }
    \label{fig:ca_trans_all}
\end{figure*}

As shown in Figure~\ref{fig:ca_trans_all}, the individual \ion{Ca}{2} H \& K lines are clearly detected despite their location near the blue end of the spectrum (3968.47~\AA\ and 3933.66~\AA, respectively).
As in the cross-correlation analysis, the \ion{Ca}{2} lines appear blue-shifted by a few \kms\ with respect to their predicted location (Table~\ref{tab:transmission_fits}).
\begin{table}
\centering
\begin{tabular}{l||rrr}
\hhline{====}
Line & Depth (\%) & $\Delta v_{\rm{sys}}$ & FWHM \\
\hline
\ion{Ca}{2} H & $ 3.25 \pm 0.37 $ &  $ -4.2 \pm 1.6 $ &  $ 28.3 \pm 3.7 $ \\
\ion{Ca}{2} K & $ 4.30 \pm 0.36 $ &  $ -4.2 \pm 0.9 $ &  $ 20.7 \pm 2.0 $ \\
H$\alpha$ & $ 1.56 \pm 0.15 $ &  $ -2.5 \pm 1.6 $ &  $ 34.1 \pm 3.9 $\\
H$\beta$ & $ 0.80 \pm 0.14 $ &  $ 6.8 \pm 2.8 $ &  $ 32.9 \pm 6.7 $ \\
H$\gamma$ & $ 1.07 \pm 0.18 $ &  $ -6.8 \pm 3.1 $ &  $ 36.5 \pm 7.2 $ \\
\ion{Na}{1} D1 & $ 0.61 \pm 0.15 $ &  $ -2.7 \pm 1.2 $ &  $ 10.1 \pm 2.9 $ \\
\ion{Na}{1} D2 & $ 0.70 \pm 0.14 $ &  $ -6.1 \pm 1.1 $ &  $ 12.3 \pm 2.6 $ \\
\hline
\end{tabular}
\caption{Results from the transmission spectroscopy analysis. For each spectral feature, we indicate the depth, velocity offset (in \kms) and FWHM (in \kms), derived from the Gaussian fit.}\label{tab:transmission_fits}
\end{table}

The \ion{Ca}{2} H \& K lines have larger absolute depths than those reported in WASP-33\,b and KELT-9\,b \citep{yan2019}. We can use Equation~\ref{eq:depth} to calculate the excess planet radii for each of the \ion{Ca}{2} lines in HAT-P-70\,b: $\Delta R_p/R_p = 1.08\pm0.09$ for the H line, and $\Delta R_p/R_p = 1.32\pm0.08$ for the K line. These values imply the presence of this ionized species at very high altitudes. \ion{Ca}{2} appears to be common in the extended atmospheres of ultra-hot Jupiters \citep{casasayasbarris2019,yan2019,turner2020,nugroho2020,tabernero2020,borsa2021,merritt2021}, where it outnumbers \ion{Ca}{1}. The H \& K lines are the dominant spectral features in the \ion{Ca}{2} high-resolution models (Figure~\ref{fig:templates_all}).

Using the planet mass upper limit of $\mpl<6.78\,\mjup$ set by \citet{zhou2019}, the Roche lobe radius of HAT-P-70\,b is $R_{\rm{Roche}}/R_p<4$ \citep{eggleton1983}. If $(R_p + \Delta R_p) > R_{\rm{Roche}}$ (i.e. if $\mpl \lesssim 1.3\,\mjup$), then that would indicate that \ion{Ca}{2} is escaping from the atmosphere.

\subsubsection{Detection of H$\alpha$, H$\beta$ and H$\gamma$}
\begin{figure*}
    \centering
    \includegraphics[width=\linewidth]{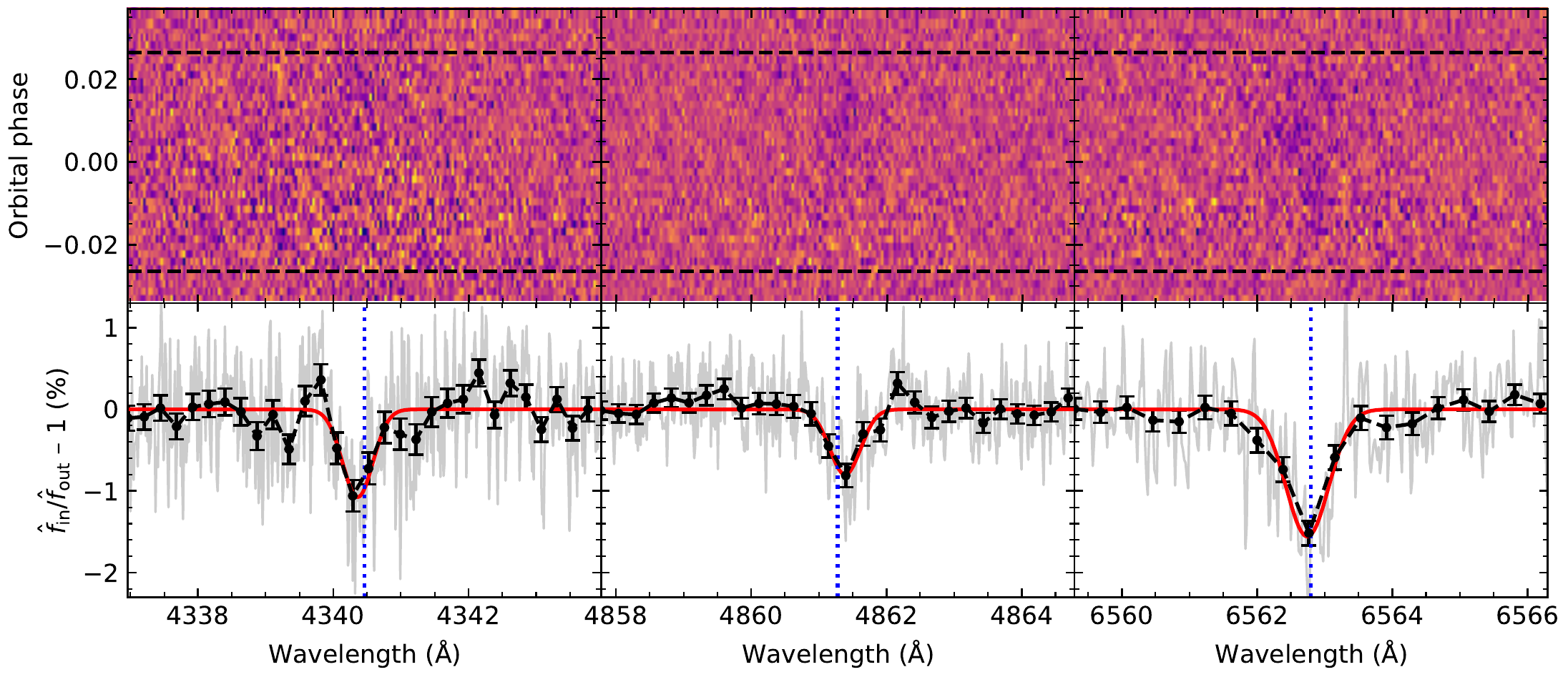}
    \caption{Same as Figure~\ref{fig:ca_trans_all}, but for H$\gamma$ (left), H$\beta$ (middle) and H$\alpha$ (right).}
    \label{fig:h_trans_all}
\end{figure*}

In addition to the species found by cross-correlation, the transmission spectroscopy analysis reveals one more species: \ion{H}{1}. In particular, we detect absorption by the H$\alpha$ (6562.80~\AA), H$\beta$ (4861.28~\AA) and H$\gamma$ (4340.46~\AA) Balmer lines (Figure~\ref{fig:h_trans_all}). Their corresponding excess transit depths ($\Delta R_p/R_p$) are $0.61\pm0.05$, $0.35\pm0.05$ and $0.45\pm0.07$, respectively.
This suggests that HAT-P-70\,b has an extended atmosphere where hydrogen atoms become excited by the extreme ultraviolet (EUV) radiation coming from the host star, as previously observed in other ultra-hot Jupiters around A-type stars, such as KELT-9\,b \citep{yan2018,cauley2019,wyttenbach2020}, MASCARA-2\,b \citep{casasayasbarris2018,casasayasbarris2019} and WASP-33\,b \citep{cauley2020,yan2021}. Excess H$\alpha$ absorption has also been reported in ultra-hot Jupiters orbiting cooler stars, such as WASP-12\,b \citep{jensen2018} and WASP-121\,b \citep{cabot2020,borsa2021}, and it can be indicative of an outflowing envelope \citep{yan2018}. The simultaneous detection of H$\alpha$, H$\beta$ and H$\gamma$ in HAT-P-70\,b offers the opportunity to constrain the temperature structure of its upper atmosphere \citep{wyttenbach2020,fossati2020} and investigate a potential Balmer-driven hydrodynamic escape \citep{garciamunoz2019}.

\subsubsection{Detection of the \ion{Na}{1} doublet}
\begin{figure*}
    \centering
    \includegraphics[width=\linewidth]{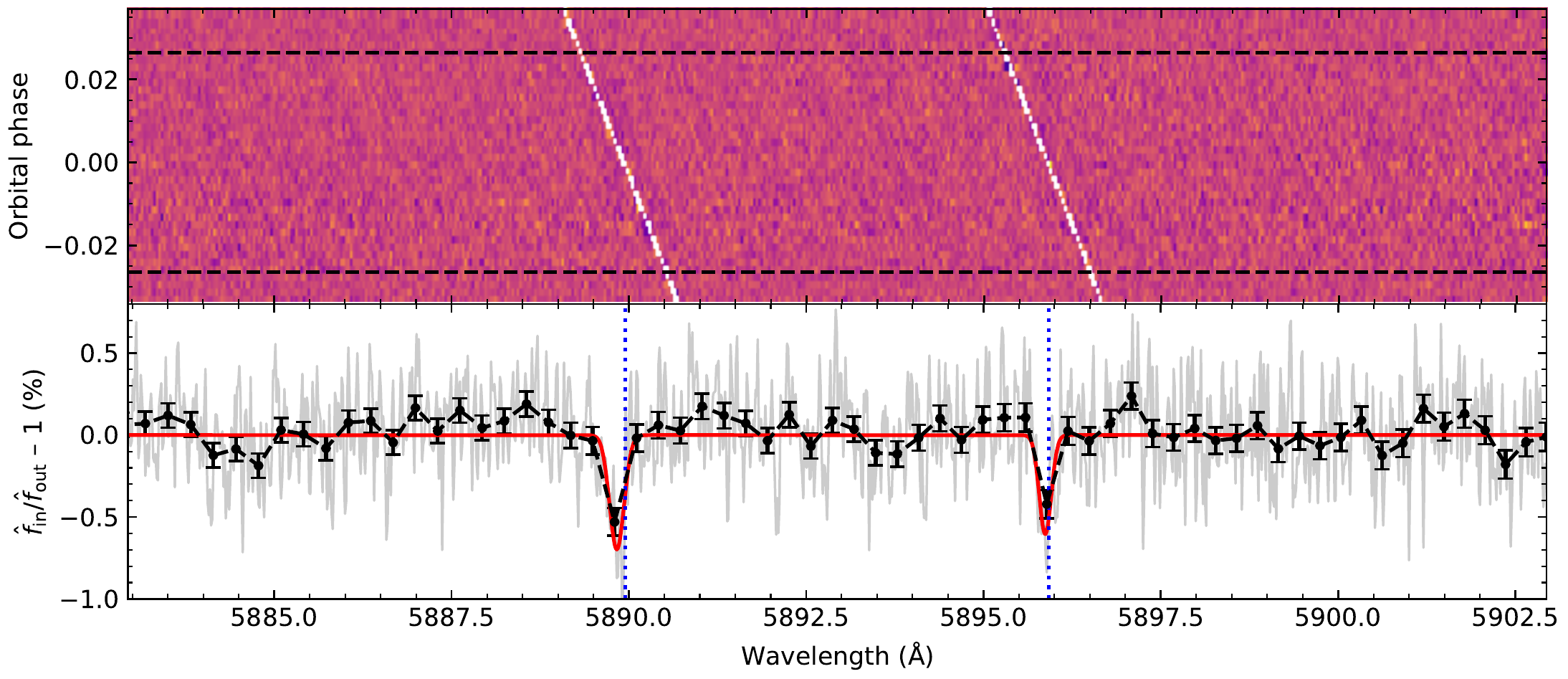}
    \caption{Same as Figure~\ref{fig:ca_trans_all}, but for the \ion{Na}{1} doublet. The atmospheric signal appears as a dark, vertical trail (top panel). The white trails are the masked out artifacts from the interstellar medium absorption lines.}
    \label{fig:na_trans_all}
\end{figure*}

In Figure~\ref{fig:na_trans_all} we show the transmission spectrum in the region of the \ion{Na}{1} doublet. We find absorption at the location of the \ion{Na}{1} D1 (5895.924~\AA) and D2 (5889.951~\AA) lines, with similar excess transit depths ($\Delta R_p / R_p = 0.27 \pm 0.06$ and $\Delta R_p / R_p = 0.31 \pm 0.06$). Both features are blue-shifted by a few \kms\ from their expected location, consistent with the results in the cross-correlation analysis.

\section{Conclusions}\label{sec:conclusions}
In this work we present the first analysis of the transmission spectrum of the ultra-hot Jupiter HAT-P-70\,b using high-resolution data from HARPS-N. Using a cross-correlation analysis, we detect eight neutral and singly ionized atomic species (\ion{Ca}{2}, \ion{Cr}{1}, \ion{Cr}{2}, \ion{Fe}{1}, \ion{Fe}{2}, \ion{Mg}{1}, \ion{Na}{1} and \ion{V}{1}) and find tentative evidence of \ion{Ca}{1} and \ion{Ti}{2}. These species appear blue-shifted by a few \kms, consistent with winds that have a predominantly subsolar-antisolar circulation pattern, similar to other ultra-hot Jupiters \citep[e.g.][]{stangret2020,nugroho2020,borsa2021}. Additionally, via transmission spectroscopy, we individually resolve the \ion{Ca}{2} H \& K lines (formed at very low pressures), the \ion{Na}{1} doublet and the H$\alpha$, H$\beta$ and H$\gamma$ lines, which indicate the presence of an extended \ion{H}{1} envelope. 

HAT-P-70\,b joins the family of ultra-hot Jupiters with a chemically rich spectrum. We emphasize that, despite the high number of species we detected in HAT-P-70\,b and the remarkable strength of some of them (e.g. \ion{Fe}{2}), the work presented here only used a single transit event with HARPS-N --- further studies could reveal additional species and allow for a detailed investigation of the atmospheric dynamics of HAT-P-70\,b \citep{wardenier2021,seidel2021}. HAT-P-70\,b is also a promising target for emission spectroscopy to investigate the chemistry of its day-side \citep[e.g.][]{pino2020,nugroho2020emission} and a possible temperature inversion \citep{lothringer2019,malik2019}.

\acknowledgments
We would like to thank the anonymous referee for a thoughtful report, which improved the quality of the manuscript. We would also like to thank René Tronsgaard for enlightening discussions preparing the HARPS-N observations, and Simon Albrecht for helpful insights on the obliquity analysis. A.B.-A. gratefully acknowledges support from "la Caixa" Foundation (ID 100010434), under agreement LCF/BQ/EU19/11710067. This work is based on observations made with the Italian Telescopio Nazionale Galileo (TNG) operated on the island of La Palma by the Fundación Galileo Galilei of the INAF (Istituto Nazionale di Astrofisica) at the Spanish Observatorio del Roque de los Muchachos of the Instituto de Astrofisica de Canarias. This work has made use of the VALD database, operated at Uppsala University, the Institute of Astronomy RAS in Moscow, and the University of Vienna.

%

\vspace{5mm}
\facilities{TNG (HARPS-North)}


\software{astropy \citep{2013A&A...558A..33A},  
\texttt{corner} \citep{corner},
\texttt{emcee} \citep{emcee},
\texttt{HELIOS-K} \citep{2021Grimm},
\texttt{matplotlib} \citep{Hunter2007},
\texttt{molecfit} \citep{smette2015,kausch2015}},
\texttt{numpy} \citep{numpy2020},
\texttt{PyAstronomy} \citep{czesla2019},
\texttt{scipy} \citep{scipy2020},
\texttt{spectres} \citep{carnall2017}.

\clearpage
\appendix

\section{Doppler shadow posteriors}
\begin{figure*}[h!]
    \centering
    \includegraphics[width=.96\linewidth]{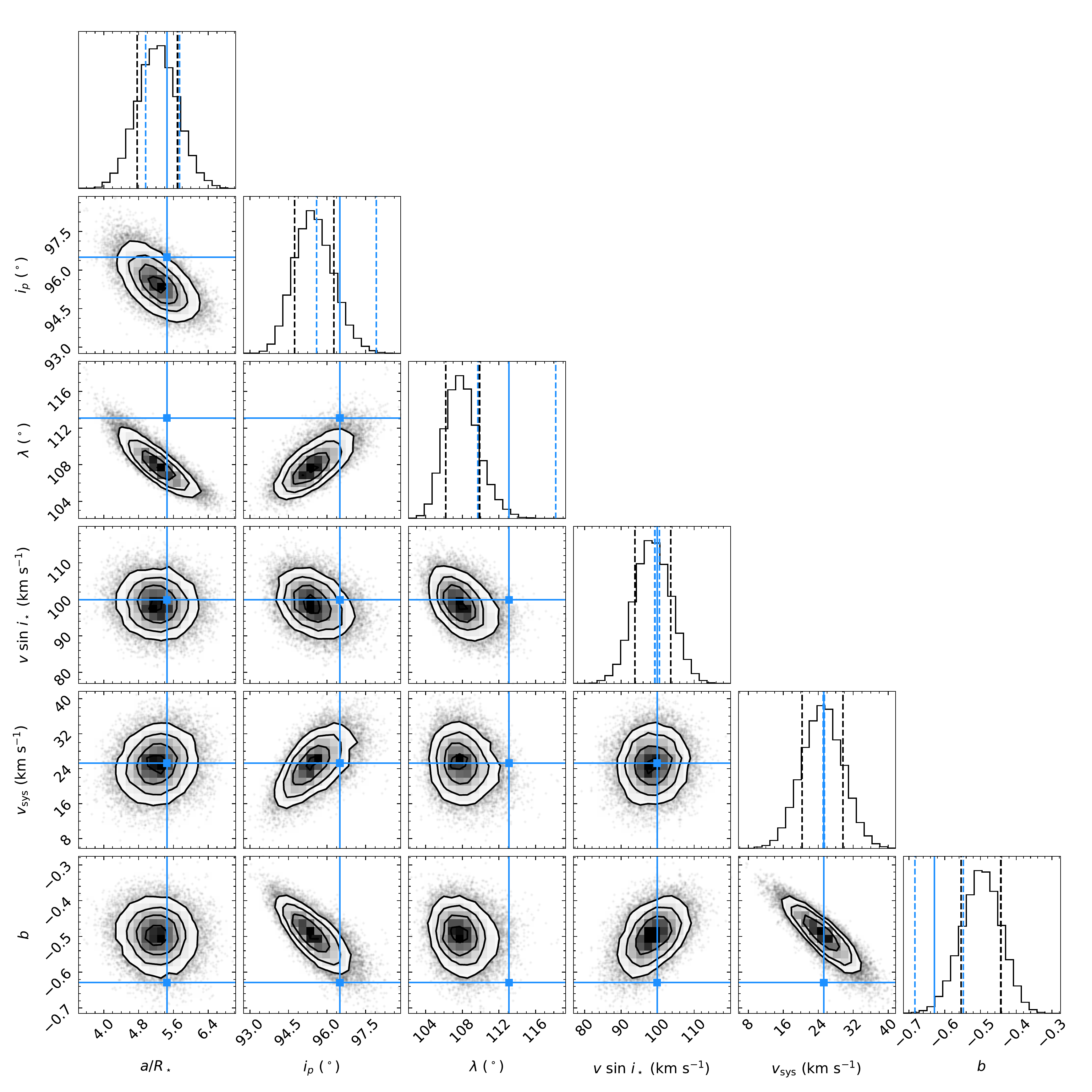}
    \caption{Corner plot of the posteriors of our MCMC analysis of the Doppler shadow. The black dashed lines in the histograms correspond to the 16th and 84th percentiles. The blue solid lines indicate the values in \citet{zhou2019}, with the corresponding uncertainty intervals marked by blue dashed lines.}
    \label{fig:posteriors}
\end{figure*}

Figure~\ref{fig:posteriors} presents the posteriors from the MCMC analysis of the Doppler shadow, used to measure the obliquity of HAT-P-70\,b. We also plot the impact parameter ($b=a\cos{i_p}/R_\star$). Although $b$ is not an independent parameter, including it in Figure~\ref{fig:posteriors} helps to visualize the correlation between different parameters.

\bibliography{sample631}{}
\bibliographystyle{aasjournal}



\end{document}